\documentclass[aip,jcp,graphicx,reprint]{revtex4-1}
\usepackage{graphicx} 
\usepackage{amsmath}

\begin{document}

\title{Multimer Embedding for Molecular Crystals Utilizing up to Tetramer Interactions}

\author{Alexander List}
\author{A. Daniel Boese}
\email[]{adrian\_daniel.boese@uni-graz.at}
\author{Johannes Hoja}
\email[]{johannes.hoja@uni-graz.at}
\affiliation{Department of Chemistry, University of Graz, Heinrichstraße 28/IV, 8010 Graz, Austria.}

\date{18 December 2025}

\begin{abstract}
Molecular crystals possess a highly complex crystallographic landscape which in many cases results in the experimental observation of multiple crystal structures for the same compound.
Accurate results can often be obtained for such systems by employing periodic density functional theory using hybrid functionals; however, this is not always computationally feasible.
One possibility to circumvent these expensive periodic calculations is the utilization of multimer embedding methods. 
Therein, the fully periodic crystal is described at a lower level of theory, and subsequently monomer energies, dimer interaction energies, etc. are corrected via high-level calculations.
In this paper, we further extend such a multimer embedding approach by one multimer order for all investigated properties, allowing us to compute lattice energies up to the tetramer embedding level, and atomic forces, the stress tensor, and harmonic phonons up to the trimer level.
We test the significance of including these higher-order multimers by embedding PBE0+MBD multimers into periodic PBE+MBD calculations utilizing the X23 benchmark set of molecular crystals and comparing the results to explicit periodic PBE0+MBD calculations.
We show that tetramer interactions systematically improve the lattice energy approximation and explore multiple possibilities for multimer selection.
Furthermore, we confirm that trimer interactions are crucial for the description of the stress tensor, yielding cell volumes within 1~\% of those of PBE0+MBD.
Subsequently, this also results in an improvement of the description of vibrational properties, giving on average gamma point frequencies within 1.3~cm$^{-1}$ and vibrational free energies within 0.3~kJ/mol of the PBE0+MBD results.
\end{abstract}


\maketitle 

\section{Introduction}
Molecular crystals constitute a crucial class of solid-state materials with applications ranging from pharmaceuticals\cite{Lee2014,Ma2023} and organic semiconductors\cite{Yang2018,Anthony2007} to explosives\cite{Ma2014}.
The molecules within such crystals are held together only by relatively weak and complex non-covalent interactions.
As a consequence, their crystal packing is highly sensitive to changes in the crystallization environment and subtle changes in the intermolecular interactions, leading to the existence of more than one distinct crystal structure in many cases.
These so-called polymorphs may exhibit vastly different physical properties yet have similar lattice energies\cite{CruzCabeza2015,Chattopadhyay2025}, often within a few kJ/mol. 
Therefore, an accurate theoretical description of these complex intermolecular interactions and a reliable prediction of all possible polymorphs -- a task generally referred to as crystal structure prediction (CSP) -- remains highly non-trivial and requires very accurate computational methods.

To track the progress in this difficult endeavor, the Cambridge Crystallographic Data Centre (CCDC) regularly organizes CSP blind tests, in which different computational methods and protocols are evaluated by their ability to accurately predict experimentally observed crystal structures.
In earlier blind tests\cite{Lommerse2000, Motherwell2002, Day2005, Day2009, Bardwell2011, Reilly2016}, success was measured in a single stage, where participants were initially provided with only the 2D Lewis structure of a molecule and had to return a list of energetically ranked 3D molecular crystal structures, hopefully containing the experimentally observed polymorphs.
In contrast, the most recent blind test was split into two stages: The first stage\cite{Hunnisett2024a} focused on the ability of generation methods to produce experimental polymorphs, where Monte Carlo methods have shown impressively high success rates.
The second stage\cite{Hunnisett2024b}, which is of higher relevance to this work, then investigated the ability of energy ranking methods to re-rank experimental crystal structures among the most stable ones.
Here, periodic density functional theory (DFT) methods have shown the best agreement with experimental data, with there being several cases where going beyond the generalized gradient approximation (GGA) and/or including thermal effects has been beneficial, for example, utilizing hybrid density functionals and the harmonic approximation for vibrational contributions.
Other examples in literature also report improvements in the description of molecular crystals when going from the GGA level, such as PBE+MBD, to a hybrid functional like PBE0+MBD\cite{Reilly2013,Reilly2013b,Marom2013,Reilly2015,Shtukenberg2017,Hoja2018,Hoja2019,Firaha2023}, or when considering thermal effects.\cite{Reilly2014,Nyman2015,Hoja2019,Firaha2023}

Unfortunately, for systems of practical relevance, such as many of those investigated in the CCDC blind tests, the system sizes and number of required calculations prohibit the extensive use of the desired level of theory.
Not only do converged periodic hybrid DFT calculations require a large amount of CPU time, they also need an exceptional amount of memory, especially when supercells are needed, which is the case for phonon calculations.

One possibility to eliminate the need of prohibitively expensive calculations is the utilization of so-called embedding schemes, in which such calculations are broken up into smaller, manageable fragments.\cite{Wen2012,Beran2016,SchmittMonreal2020,Paulus2006,Hermann2009,Bates2011,Bates2011b,Pruitt2012,Herbert2019,Liu2019}
In the case of molecular crystals, this fragmentation usually involves the crystal's individual monomers, dimers, trimers, etc.
Generally, there are two types of multimer embedding schemes to be distinguished:
First, there is the additive scheme\cite{Yang2014,Cervinka2016,Teuteberg2019,Nelson2024,Pham2023,Pham2024}, in which a crystal's monomer energies, dimer interaction energies, trimer interaction energies, etc., are summed up. 
This expansion will eventually converge towards the energy of the explicit periodic calculation without having ever performed such.
For this additive scheme, up to tetramer interactions are regularly needed, as they will still contribute a significant part to the lattice energy.
However, the required distance cutoffs are generally large, stretching to several tens of Angstroms, resulting in thousands of dimers, trimers, and tetramers to be calculated.
Because of this, most of these studies only consider energies.

The other possibility is to employ a subtractive embedding scheme\cite{Beran2010,Wen2011,Nanda2012,Boese2017,Loboda2018,Dolgonos2018,Hoja2023,Syty2025}.
Here, a computationally cheaper lower-level method other than the desired one is used to carry out a canonical periodic molecular crystal calculation.
Monomers, dimers, trimers, etc. within the crystal are then evaluated at both the high- and low-level method and the low-level periodic calculation is corrected by adding the difference between the two methods for all multimer contributions.
The benefit of this approach is that many-body effects up to infinite order as well as long-range effects are already included via the periodic low-level calculation.
Thus, the many-body expansion of the embedding scheme can be truncated earlier and with smaller cutoff distances than in the additive case, especially when the low- and high-level methods are chosen to be compatible.
We have previously developed such a subtractive embedding scheme\cite{Loboda2018, Hoja2023} and have also applied it to the second stage of the most recent CCDC blind test\cite{Hunnisett2024b}, where we embedded PBE0+MBD monomers and dimers into periodic PBE+MBD calculations for lattice energies, geometry optimizations, and partly for phonon calculations.
Using this approach, we were able to obtain the most stable experimental crystal structure at rank 1 in our stability ranking for 4 out of 5 systems.

In this paper, we further extend the above-mentioned embedding approach.
Compared to our previous publication\cite{Hoja2023}, we have added one additional multimer order to the calculation of each investigated property, being now able to determine embedded lattice energies up to the tetramer level, and atomic forces, the stress tensor, and harmonic phonons up to the trimer level.
The performance of these new methods is evaluated by embedding PBE0+MBD multimers into periodic PBE+MBD calculations and comparing the results to those of explicit periodic PBE0+MBD calculations utilizing the X23\cite{OterodelaRoza2012,Reilly2013,Reilly2013b,Dolgonos2019} benchmark set of molecular crystals, a well-established standard for testing and developing methods for molecular crystals.\cite{Moellmann2014,Ikabata2014,Grimme2015,Cutini2016,Nyman2016,Gould2016,Mortazavi2018,Stein2019,Chen2020,Gale2021,Jana2021,Tuca2022,Price2023,DellaPia2024}
Additionally, for lattice energies, we investigate the importance of the different tri- and tetramer types, which arise when applying finite distance cutoffs for multimer selection, the possibility of applying different distance cutoff values for each multimer order, and the application of energy-based cutoffs to increase the efficiency of our embedding approach.

\section{Computational Methods}

\subsection{Energy}
The high-level periodic energy of a molecular crystal, $E_\mathrm{per}^\mathrm{high}$, is approximated in our subtractive multimer embedding approach as
\begin{equation}
\begin{split}
     \label{eq:energy}
    E_\mathrm{per}^\mathrm{high} \approx E_\mathrm{per}^\mathrm{low} 
    + \sum_i n_i \Delta E_i
    + \sum_{i>j} \frac{n_{ij}}{2} \Delta E_{ij}^\mathrm{int} \\
    + \sum_{i>j>k} \frac{n_{ijk}}{3} \Delta E_{ijk}^\mathrm{int}
    + \sum_{i>j>k>l} \frac{n_{ijkl}}{4} \Delta E_{ijkl}^\mathrm{int},
\end{split}
\end{equation}
where the final term has now been added to account for tetramer interaction energies $E_{ijkl}^\mathrm{int}$ as well, with a tetramer $ijkl$ consisting of the monomers $i$, $j$, $k$, and $l$.
These tetramer interaction energies are defined as the tetramer energy $E_{ijkl}$, from which all trimer interaction energies $E_{xyz}^\mathrm{int}$, dimer interaction energies $E_{xy}^\mathrm{int}$, and monomer energies $E_{x}$ are deducted ($\{x\}, \{x,y\}, \{x,y,z\} \subset \{i,j,k,l\}$):
\begin{equation}
\begin{split}
    E_{ijkl}^\mathrm{int} = E_{ijkl}
    - E_{ijk}^\mathrm{int} - E_{ijl}^\mathrm{int}
    - E_{ikl}^\mathrm{int} - E_{jkl}^\mathrm{int} \\
    - E_{ij}^\mathrm{int} - E_{ik}^\mathrm{int}
    - E_{il}^\mathrm{int} - E_{jk}^\mathrm{int} 
    - E_{jl}^\mathrm{int} - E_{kl}^\mathrm{int} \\
    - E_{i} - E_{j} - E_{k} - E_{l}
\end{split}
\end{equation}
Trimer and dimer interaction energies are defined analogously and have been explicitly described in our previous publication \cite{Hoja2023}.
For each multimer contribution, the respective $n$ amounts to how many monomers of the multimer belong to the main unit cell, \emph{i.e.}, the central unit cell in the supercell from which the multimers are extracted.
In the context of tetramers, their interaction energy is counted fully, three quarters, two quarters, one quarter, or not at all if four, three, two, one, or none of their monomers belong to the main unit cell, respectively.

Any $\Delta E$ term refers to the difference between the high- and low-level energy of the system:
\begin{equation}
    \Delta E = E_\mathrm{high} - E_\mathrm{low}.
\end{equation}
The lattice energy $E_\mathrm{latt}^\mathrm{high}$ is given by
\begin{equation}
    E^{\rm high}_{\rm latt} = \frac{E_\mathrm{per}^\mathrm{high}}{Z} - E_\mathrm{mon}^\mathrm{high},
\end{equation}
where $Z$ is the number of molecules in the unit cell of the system and $E_\mathrm{mon}^\mathrm{high}$ is the high-level energy of an isolated monomer of the system in its lowest-energy gas-phase conformation.

\begin{figure}[!]
\includegraphics[width=\columnwidth]{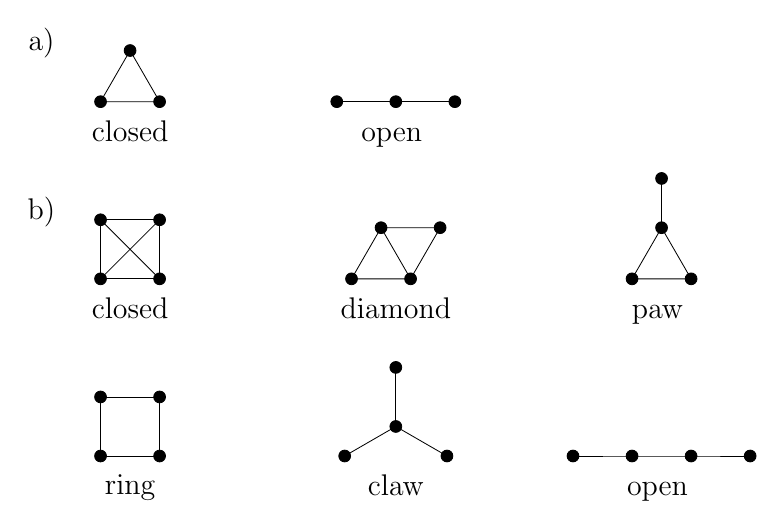}
\caption{\label{fig:multimer_types} Different types of a) trimers and b) tetramers. Each vertex represents a monomer, and edges indicate whether or not two monomers are within cutoff distance.}
\end{figure}

Previously, for trimers to be considered, the intermolecular distance between all three monomer pairs had to be smaller than the defined multimer cutoff distance.
From now on, we will refer to these trimers as 'closed' type.
Herein, we also investigate the other type of trimer, for which only two of the three monomer pairs are within a cutoff distance, named the 'open' type.
Going to tetramers, there are more types to differentiate.
If the four contributing monomers are thought of as vertices, and edges between the vertices indicate an intermolecular distance lower than the multimer cutoff distance, one obtains six non-isomorphic connected graphs.
Therefore, one can differentiate between six unique tetramer types.
We name them either by borrowing names from graph theory if they are descriptive or introducing new names.
In descending order of edges (pairs of monomers within cutoff distance), these are: closed (all 6 edges), diamond (5 edges), paw (4 edges), ring (4 edges), claw (3 edges), and open (3 edges).
Visuals for the tri- and tetramer types are provided in Fig. \ref{fig:multimer_types}.
As per definition, non-closed multimers contain lower-order multimers which are not recognized as such due to their intermolecular distances being larger than the cutoff value.
In such cases, the interaction energy of these missing lower-order multimers is set to zero.

Calculations utilizing only closed multimers will keep the original label, namely ME$X$($x$\AA{})(PBE0+MBD:PBE+MBD), where $X$ specifies the utilized multimer order, \emph{i.e.} 1 for only monomers, 2 for monomers and dimers, 3 for up to trimers, and 4 for up to tetramers, and $x$\AA{} defines the multimer cutoff distance.
Since in this paper we only embed PBE0+MBD into PBE+MBD, we will from now on omit the information in the latter parenthesis.
Calculations where all types of multimers up to the respective level are utilized will be marked with an 'a' in front of the cutoff distance, the label then being ME$X$(a$x$\AA{}).
For tetramers, where in-between cases are possible, additional labels are introduced to indicate which tetramer types are included, specifics of which are provided in the description of Fig. S1 in the supplementary material.
For all of those in-between cases of tetramer type inclusion, open trimers are included as well.

Furthermore, we consider the possibility of utilizing mixed cutoff distances for the multimers of different orders.
For trimer embedding, the label is then ME$3$($x$/$y$\AA{}), where $x$ and $y$ are the respective dimer and trimer cutoff distances.
For tetramer embedding, the label is ME$4$($x$/$y$/$z$\AA{}), with $x$, $y$, and $z$ being the dimer, trimer, and tetramer cutoff distances, respectively.
From a methodological standpoint, it is only sensible to decrease the cutoff when going to higher multimer orders, as otherwise, energies of lower-order multimers would be absent when calculating higher-order multimer interaction energies.

Lastly, in an attempt to decrease the number of multimers that have to be calculated, we introduce energy-based cutoffs in addition to the distance-based cutoff.
For this approach, all low-level calculations are performed first, and for the highest-order multimers, only those whose low-level interaction energy exceeds a threshold are considered further, saving the computational cost of the high-level calculation of the others.
We label such calculations as ME$X$($x$\AA{} $>n$~kJ/mol), indicating that only the highest-order multimers whose low-level interaction energy magnitude exceeds $n$~kJ/mol are included.

\subsection{Forces}
The high-level forces $\mathbf{f}_\mathrm{per}^\mathrm{high}(a)$ of an atom $a$, contained within monomer $i$ and belonging to the main unit cell, are approximated as
\begin{equation}
    \mathbf{f}_\mathrm{per}^\mathrm{high}(a) \approx
    \mathbf{f}_\mathrm{per}^\mathrm{low}(a)
    +  \Delta \mathbf{f}_i(a)
    + \sum_{j} \Delta \mathbf{f}_{ij}^\mathrm{int}(a)
    + \sum_{j>k} \Delta \mathbf{f}_{ijk}^\mathrm{int}(a) \, ,
\end{equation}
which, compared to the previous formulation, have been extended by the final term that sums up the high- and low-level differences between all the trimer interaction forces experienced by atom $a$ ($i \neq j, i\neq k)$:
 \begin{equation}
    \mathbf{f}_{ijk}^\mathrm{int}(a) =
    \mathbf{f}_{ijk}(a)
    - \mathbf{f}_{ij}^\mathrm{int}(a) - \mathbf{f}_{ik}^\mathrm{int}(a)
    - \mathbf{f}_{i}(a).
\end{equation}

\subsection{Stress}
The stress tensor components $\sigma_{pq}^\mathrm{high}$ ($p,q \in \{1, 2, 3\}$) of the unit cell at the high-level periodic method are approximated within our embedding scheme as
\begin{equation}
\begin{split}
    \sigma_{pq}^\mathrm{high} \approx \sigma_{pq}^\mathrm{low} 
    -\frac{1}{V} 
    \sum_i  \sum_{a} n_i r_{i,p}(a)\, \Delta f_{i,q}(a)\\
    -  \frac{1}{V}  \sum_{i>j}  \sum_{a} \frac{n_{ij}}{2} r_{ij,p}(a) \, \Delta f_{ij,q}^\mathrm{int}(a)\\
    -  \frac{1}{V}  \sum_{i>j>k}  \sum_{a} \frac{n_{ijk}}{3} r_{ijk,p}(a) \, \Delta f_{ijk,q}^\mathrm{int}(a) \, ,
\end{split}
\end{equation}
which has also been extended by the final term to account for trimer interactions.
$\sigma_{pq}^\mathrm{low}$ is the respective component of the low-level stress tensor, and $V$ is the unit cell volume.
The first summation in all the other terms sums up over all monomers, dimers, and trimers, respectively, while the second summation sums up over all atoms $a$ contained within the respective multimer.
The terms which are summed up are the products of the $x$, $y$, or $z$ component of the position $\mathbf{r}$ of atom $a$ and the $x$, $y$, or $z$ component of the high- and low-level force $\mathbf{f}$ or force interaction $\mathbf{f}^\mathrm{int}$ difference of the respective multimer acting on atom $a$, again under the consideration of how many monomers of the multimer are contained within the main unit cell.

\subsection{Harmonic Vibrational Properties}
The approach to calculate harmonic vibrational properties remains the same as in Ref. \citenum{Hoja2023}, except now up to trimers are included.
Utilizing phonopy \cite{Togo2015}, finite displacements within sufficiently large supercells are created and their low-level atomic forces calculated.
For all multimers that contain an atom which was displaced, the same displacement is applied to them as well and they are recalculated at the low and high level to correct the periodic low-level forces of the phonopy supercell by the difference in the same fashion as described in section B.
The resulting approximated high-level force sets are then used to obtain the vibrational properties via phonopy.

\subsection{Computational Details}
All multimer embedding calculations were carried out using our open-source code MEmbed \cite{membed} (version 0.3.0), which in turn utilizes the Atomic Simulation Environment \cite{HjorthLarsen2017} (ASE) for several of its functionalities.
The electronic structure calculations were carried out using FHI-aims \cite{Blum2009,Knuth2015,Ren2012,Yu2018,Havu2009,Ihrig2015} (version 231212) via the interface provided by ASE.
For the low- and high-level method, the PBE \cite{Perdew1996} and PBE0 \cite{Adamo1999} density functional approximations were used, respectively, together with the many-body dispersion \cite{Tkatchenko2012,Ambrosetti2014} (MBD) method (rsSCS version) to correct the van-der-Waals dispersion interactions.
Throughout, we utilize light species default settings within FHI-aims for integration grids and basis functions (2020 version) for both periodic and isolated systems, being able to do so due to the employed numeric atom-centered basis functions.
In this paper, we limit ourselves to using light settings for computational efficiency, as we have already established in the previous publication \cite{Hoja2023} that our embedding scheme behaves similarly for tight settings as it does for light settings.

For the single point calculations, the total energy, the forces, the charge density, and the sum of eigenvalues were converged to $10^{-6}$~eV, $10^{-4}$~eV/\AA{}, $10^{-5}$~electrons/\AA{}$^3$, and $10^{-3}$~eV, respectively, for both periodic as well as low- and high-level multimer calculations.
For the periodic DFT calculations, the k-grids were chosen to satisfy $nx > 18$~\AA{}, with $x$ being the cell length in the respective direction and $n$ being the number of k-points in that direction.
For MBD energies and forces, the k-grid was chosen to satisfy $nx > 25$~\AA{}.
To calculate lattice energies, we used the energies of the isolated molecules from our previous publication \cite{Hoja2023}.
Lattice relaxations were converged to a maximum force of $0.005$~eV/\AA{} using ASE with the BFGS algorithm and the FrechetCellFilter class while keeping the structures symmetrized during optimizations using the FixSymmetry class.
As we have switched to newer versions for all software MEmbed depends on, we carried out re-optimizations for a selection of previously investigated methods, which we use here to add perspective to newly obtained results.
In almost all cases, this resulted in no changes, except for 4 of the 23 PBE0+MBD/light-optimized structures, which required additional optimization steps.
Note that these changes have no impact on any previous findings; nonetheless, for these previously investigated methods, we have provided the new lattice energies in Table S7 in the supplementary material and the corresponding error metrics in Table \ref{tbl:energies_closed}.
Phonon calculations were performed on supercells generated from the respective optimized geometries so that each cell length exceeded $12$~\AA{},
utilizing finite displacements of $0.005$~\AA{}.
To evaluate the vibrational free energies, q-grids were chosen to satisfy $nx > 50$~\AA{}, resulting in all acoustic modes being smaller than 0.5~cm$^{-1}$ in magnitude at the gamma point.

\section{Results and Discussion}

\subsection{Lattice Energies for up to Tetramer Embedding}

\begin{figure*}[!]
\includegraphics[width=\textwidth]{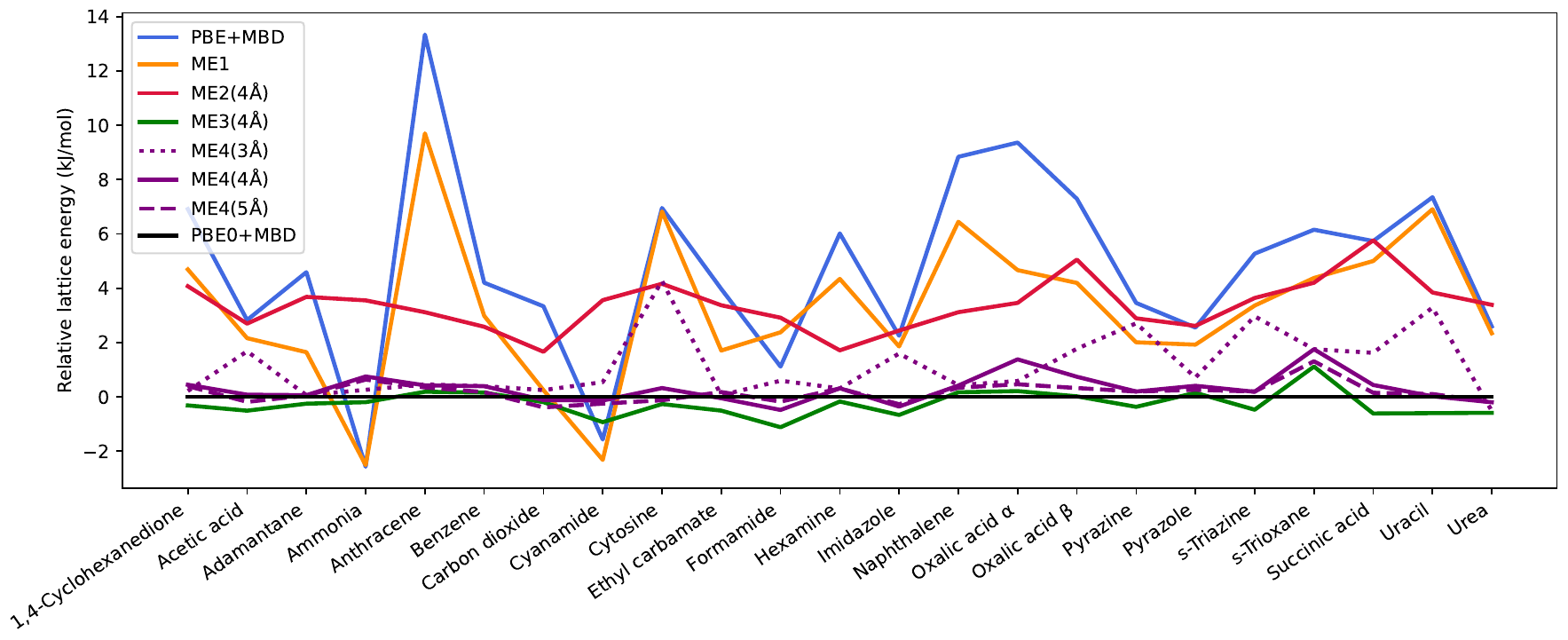}
\caption{\label{fig:energy_overview} Relative lattice energies in kJ/mol of the X23 set calculated with several approaches w.r.t PBE0+MBD. All calculations were carried out on top of the PBE0+MBD/light optimized structures.}
\end{figure*}

We begin by discussing the impact of tetramer inclusion in our embedding approach on the calculation of lattice energies, considering only closed multimers for now.
Therefore, we have carried out single-point ME4($x$\AA{}) calculations with light species default settings on top of the PBE0+MBD/light-optimized X23 crystal structures, utilizing cutoff distances of $x$ = 3, 4, 5, and 6~\AA{}.
The number of unique tetramers that need to be calculated for each of the X23 systems at the various cutoff distances is provided in Table S3 in the supplementary material.
Notably, at a cutoff distance of 3~\AA{}, only 11 of the 23 benchmark systems contain closed tetramers.
The number then drastically increases with higher cutoff distances, lying between 1 and 11 at 4~\AA{}, 2 and 49 at 5~\AA{}, and 20 and 377 at 6~\AA{}.
Since our earlier publication\cite{Hoja2023}, we have extended our method to identify multimers that are identical through symmetry operations to also include improper rotation, which has reduced the number of multimers that have to be calculated explicitly for systems containing such symmetries, sometimes by up to one half.
For this reason, the new number of uniquely identified dimers and trimers is again provided here in Tables S1 and S2 in the supplementary material, respectively.

\begin{table}
\centering
\setlength{\tabcolsep}{2pt}
\caption{\label{tbl:energies_closed} Errors of the calculated lattice energies of the X23 set compared to PBE0+MBD. The tri- and tetramer embedding methods shown here utilize only multimers of the closed type. All calculations were done with light settings on top of the PBE0+MBD-optimized structures. The mean error (ME), the mean absolute error (MAE), and the maximal error (MAX) are given in kJ/mol while the mean relative error (MRE), the mean absolute relative error (MARE), and the maximal relative error (RMAX) are given in \%.}
\begin{tabular}{@{\extracolsep{4pt}}lrrrrrr@{}}
\hline\hline
Method & ME & MAE & MAX & MRE & MARE & RMAX\\
\hline
PBE+MBD              &  4.8 &  5.1 & 13.3 & -5.1 &  5.8 & 11.9 \\
ME1                  &  3.3 &  3.7 &  9.7 & -3.1 &  3.8 &  7.8 \\
ME2(4\AA{})          &  3.4 &  3.4 &  5.8 & -3.9 &  3.9 &  8.3 \\
ME3(4\AA{})          & -0.2 &  0.4 &  1.1 &  0.3 &  0.5 &  1.6 \\
ME4(3\AA{})          &  1.1 &  1.2 &  4.3 & -1.2 &  1.3 &  4.6 \\
ME4(4\AA{})          &  0.3 &  0.4 &  1.8 & -0.4 &  0.5 &  2.5 \\
ME4(5\AA{})          &  0.2 &  0.3 &  1.3 & -0.2 &  0.4 &  1.8 \\
ME4(6\AA{})          &  0.2 &  0.3 &  1.3 & -0.3 &  0.3 &  1.9 \\
\hline\hline
\end{tabular}\\
\end{table}

Fig. \ref{fig:energy_overview} shows the resulting ME4($x$\AA{}) lattice energies (the 6~\AA{} cutoff results have been left out for visual clarity), together with those of only the low-level periodic calculation (PBE+MBD/light) and a selection of previously investigated lower-order embedding methods \cite{Hoja2023} for reference.
All lattice energies are given relative to the high-level method PBE0+MBD/light.
Several error statistics are provided in Table \ref{tbl:energies_closed}, and all individual lattice energies can be found in Table S7 in the supplementary material.
Note that the positive direction in the plot and positive mean errors (ME) correspond to smaller lattice energies in magnitude, since lattice energies are negative in sign.
It can be seen that the tetramer embedding methods with cutoffs of 4~\AA{} and higher show excellent agreement with the PBE0+MBD lattice energies, in a similar manner as the best trimer embedding method, ME3(4\AA{}), which had a ME of -0.2~kJ/mol and a mean absolute error (MAE) of 0.4~kJ/mol per molecule.
With ME4(4\AA{}), the ME and MAE are 0.3~kJ/mol and 0.4~kJ/mol, respectively.
At this cutoff, all ME4 lattice energies lie above those of ME3.
We have already suggested previously that in many cases, an increasing multimer order appears to converge in the fashion of a damped oscillation, which is further implied by these results.
ME4(3\AA{}), on the other hand, leads to a ME of 1.1~kJ/mol and a MAE of 1.2~kJ/mol.
In fact, the results are very similar to those of ME3(3\AA{}), as for more than half of the systems there are no tetramers at this cutoff.

\begin{figure}[!]
\includegraphics[width=\columnwidth]{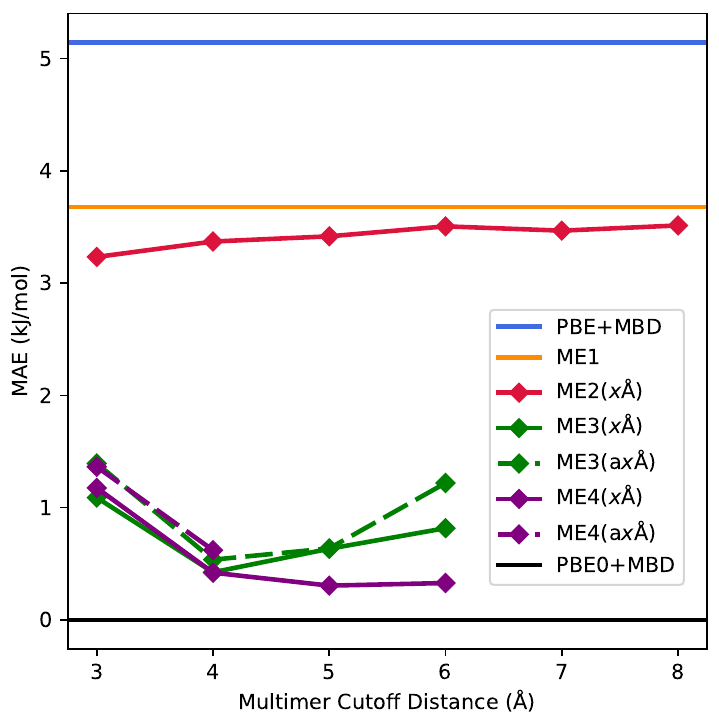}
\caption{\label{fig:energy_summary} Mean absolute errors (MAEs) of lattice energies of the X23 set as a function of the employed multimer cutoff distance for several approaches w.r.t. PBE0+MBD in kJ/mol.}
\end{figure}

For ME4(5\AA{}) and ME4(6\AA{}), the errors further decrease to a ME of 0.2~kJ/mol and a MAE of 0.3~kJ/mol in both cases.
This behavior is different from that of ME3, since there the cutoff distances larger than 4~\AA{} yielded worse results.
These circumstances are further illustrated in Fig. \ref{fig:energy_summary}, where the MAEs of the multimer embedding methods up to different orders are plotted as a function of the cutoff distance.
Whereas for trimer embedding there was a certain reliance on error cancellation to yield the results as good as they were, for tetramer embedding this reliance is much less pronounced, presumably only leading to increased errors at even larger cutoff values.
It is therefore reasonable to assume that for multimer orders beyond 4 (tetramers), a similar trend would manifest, allowing one to go to larger and larger cutoff values, eventually arriving at the high-level result.
Even though this consideration serves as a proof of concept, it is of course not practical, as the vast number of multimers to calculate would defeat the purpose of our embedding scheme to reduce computational resources.
ME3(4\AA{}) provides very good results for its expense, which can be improved upon by including tetramers at a higher cutoff value, if needed.

Next, we discuss the impact of the inclusion of tri- and tetramers which are not only of the closed type, i.e., those where not all monomer pairs are within the given cutoff distance, on the lattice energies.
Starting at the trimer level, we have performed ME3(a$x$\AA{}) single point calculations on top of the PBE0+MBD/light-optimized X23 structures, utilizing light species default settings and cutoff distances between 3 and 6~\AA{}.
These calculations are noticeably more expensive than closed-trimer calculations, as open trimers are far more numerous, as shown in Table S2 in the supplementary material.
At a 4~\AA{} cutoff distance, for example, the systems with the lowest and highest amount of unique trimers are hexamine and pyrazole, having 1 and 20 closed trimers, respectively.
When considering all trimers, an additional respective 7 and 96 open trimers have to be calculated for those systems.
The obtained lattice energies are shown in Table S8, and the respective error statistics are presented in Table S4.
The MAE for this method as a function of the cutoff distance is also shown in Fig. \ref{fig:energy_summary}.
It can be seen that overall, the inclusion of open trimer interactions does not lead to an improvement in the approximated lattice energies.
At a 3~\AA{} cutoff, the MAE is 1.4~kJ/mol, which decreases to 0.5~kJ/mol at 4~\AA{}, before then increasing again to 0.6 and 1.2~kJ/mol at 5 and 6~\AA{}, respectively, which is either identical to or up to 0.4~kJ/mol higher than for the closed-trimer calculation at the respective cutoff distance.

\begin{figure}[!]
\includegraphics[width=\columnwidth]{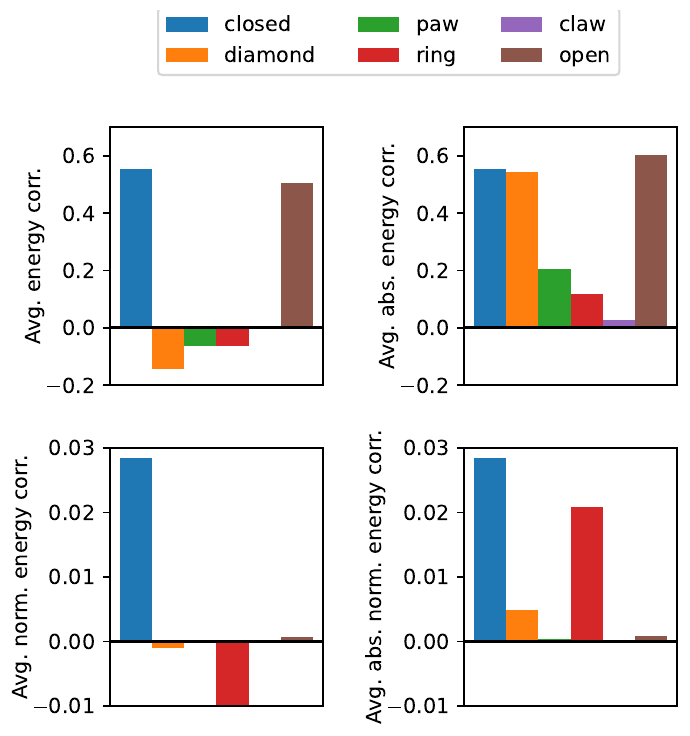}
\caption{\label{fig:tetramer_interactions} Average tetramer interaction energy correction terms in kJ/mol for the X23 crystal structures calculated with ME4(a4\AA{}) using light species default settings on top of the PBE0+MBD/light-optimized geometries, broken down to the individual tetramer types. Top left: Average of energy correction terms. Top right: Average of absolutes of energy correction terms. Bottom left: Average of energy correction terms, normalized per multimer. Bottom right: Average of absolutes of energy correction terms, normalized per multimer.}
\end{figure}

To investigate the impact of the inclusion of different tetramer multimer types on lattice energies, we have performed ME4(a$4$\AA{}) calculations, again on top of PBE0+MBD/light-optimized geometries, from which we can then extract the effect of each individual multimer type.
The number of unique occurring tetramer types varies vastly, as can be seen in Table S3.
For the X23 set at a 4~\AA{} cutoff, there are 1 to 11 unique closed tetramers, 4 to 68 diamond tetramers, 11 to 308 paw tetramers, only 1 to 7 ring tetramers, 11 to 104 claw tetramers, and 28 to 700 open tetramers.
The individual tetramer interaction correction terms for each tetramer type are shown in Fig. \ref{fig:tetramer_interactions}.
The top sub-figures show the average (left) and average of absolutes (right) correction terms, while the ones on the bottom show the same terms, but normalized per multimer.
It can be seen that the largest contributions stem from the closed, diamond, and open tetramer types, with the closed and open ones almost always having positive contributions, and the diamond tetramer contributions being roughly equally positive and negative.
Out of these three, closed tetramers have by far the largest contribution per multimer.
Interestingly, ring multimers also have a rather large contribution if normalized per multimer, but due to the fact that there are so few present, they do not make up a meaningful part of the total correction term.
Open tetramers, on the other hand, have a very small contribution per multimer, but due to there being so many of them, make up the largest portion of the total contribution.
However, as the same cutoff distance for all multimer orders is applied, two of the four trimer interactions and three of the six dimer interactions within the open tetramers do not satisfy the cutoff criterion, meaning that their interaction energies are not explicitly calculated, but instead set to zero.
We therefore presume that the unaccounted lower-order multimers, even though their contribution should be smaller than the accounted ones, artificially increase the tetramer interaction energy, which is then multiplied by the large number of open tetramers.

We now investigate the resulting lattice energies when applying the different possibilities of which tetramer types to include into ME4 calculations.
The corresponding error statistics are shown in Table S4 in the supplementary material with the individual lattice energies being presented in Table S9.
For reference, the ME4 lattice energies using only closed tri- and tetramers reach a ME of 0.3~kJ/mol, a MAE of 0.4~kJ/mol, and a MAX of 1.8~kJ/mol.
By also taking diamond tetramer interactions into account, these errors decrease to 0.1~kJ/mol, 0.3~kJ/mol, and 1.1~kJ/mol, respectively, a slightly higher improvement than the one that is observed when increasing the distance cutoff to 5~\AA{} for only closed tetramers.
Further extending the included tetramer types by paw, ring, and claw tetramers also only has a small effect, with the ME being $<$~0.1~kJ/mol for all investigated combinations, the MAE lying between 0.3 and 0.4~kJ/mol, and the maximum absolute error (MAX) between 1.2 and 1.8~kJ/mol.
While the absolute errors of the lattice energies do not improve significantly, the inclusion of more multimer types overall lowers the obtained lattice energies so that they are over- and underestimated equally as often, compared to the PBE0+MBD results.
On the other hand, after the inclusion of the final tetramer type, namely the open tetramers, we can immediately see a deterioration of the quality of obtained lattice energies, with the ME, MAE, and MAX increasing to 0.5~kJ/mol, 0.6~kJ/mol, and 2.3~kJ/mol, respectively.
Clearly, the contribution of open tetramers is overestimated, most likely due to the above-mentioned fact that half of the di- and trimer interactions are not deducted from the tetramer interactions for this tetramer type.
Overall, it seems that the inclusion of any tetramer types beyond the closed and diamond types is unnecessary.
Even then, a similar improvement can be achieved by increasing the cutoff distance to 5~\AA{} and including only closed multimers, rather than including tetramers which are considered of the diamond type at 4~\AA{}.
This is accompanied by the fact that the number of additional multimers that have to be calculated is significantly smaller for closed multimers between 4 and 5~\AA{}, compared to diamond tetramers and open trimers at up to 4~\AA{}.

\subsection{Mixed Cutoffs and Energy Cutoffs for Multimer Selection}
So far, we have only considered identical cutoff distance values for all multimer orders.
We now consider the possibility of applying mixed cutoffs to our multimer embedding scheme.
Therefore, we have performed ME3($x$/$y$) and ME4($x$/$y$/$z$) calculations using various combinations of distance cutoffs for dimers, trimers, and tetramers.
At the trimer embedding level, we have applied 3, 4, and 5~\AA{} trimer cutoffs combined with higher dimer cutoffs between 4 and 8~\AA{}.
Table S5 in the supplementary material shows that the mixed cutoff trimer embedding approaches generally reproduce the lattice energy errors of the respective calculations where the trimer cutoff is applied to both di- and trimers.
Individual lattice energies are shown in Table S10.
So for a 3~\AA{} trimer cutoff, the MEs and MAEs vary between 1.1 and 1.3~kJ/mol depending on the dimer cutoff value, for a 4~\AA{} trimer cutoff, the MEs vary between -0.2 and -0.1~kJ/mol and the MAEs between 0.3 and 0.4~kJ/mol, and for a 5~\AA{} trimer cutoff, the MEs are -0.5~kJ/mol and the MAEs lie between 0.5 and 0.6~kJ/mol.
Of the combinations tested, the one that yields the lowest overall errors is ME3(7/4\AA{}) with a ME of -0.2~kJ/mol, a MAE of 0.3~kJ/mol, and a maximum error (MAX) of 1.0~kJ/mol, which, however, is only up to 0.1~kJ/mol less than ME3(4\AA{}).
At the tetramer embedding level, combinations of di-, tri-, and tetramer cutoffs between 4 and 6~\AA{} were tested.
The resulting error statistics are also shown in Table S5 in the supplementary material and the individual lattice energies in Table S11.
Here, while the different methods show some variance in their MEs, namely between -0.2 and 0.4~kJ/mol, the MAEs are all very similar, ranging only between 0.3 and 0.5~kJ/mol.
Overall, it can be said that there is no discernible benefit from applying mixed cutoff values.

Lastly, for lattice energies, we discuss the possibility of applying an energy-based cutoff for multimer selection, in addition to the distance-based cutoff.
For this approach, all multimers are first evaluated with the low-level method, and the high-level calculations of a multimer are only carried out if the corresponding low-level interaction energy exceeds a certain threshold, the result being potential savings in the number of multimers having to be evaluated at the high-level method.
This threshold is only applied to the highest-order multimers present in the respective calculation, as otherwise, lower-order multimer energies that are needed for higher-order multimer interaction energies could be missing.
To investigate the efficiency as well as the potential loss in accuracy of such an approach, we have applied multiple different energy thresholds to the best-performing tri- and tetramer embedding methods, namely ME3(4\AA{}) and ME4(5\AA{}).
Table S6 in the supplementary material shows the resulting lattice energy errors compared to PBE0+MBD, with the individual lattice energies shown in Tables S12 and S13.
For ME3(4\AA{}), it can be seen that up to an energy threshold of 0.2~kJ/mol, the mean absolute error (MAE) and maximum error (MAX) stay constant at 0.4 and 1.1~kJ/mol, respectively, while the mean error (ME) goes from -0.2 to less than 0.1~kJ/mol.
At this threshold of 0.2~kJ/mol, the number of trimers excluded lies between 0 and 37.5~\%, with the average savings being 10~\%.
When going to a 0.5~kJ/mol energy threshold, the MAE only grows by 0.1~kJ/mol to 0.5~kJ/mol, but the ME is now 0.3~kJ/mol and the MAX grows to 1.8~kJ/mol, showing that for some systems, too important trimers are being neglected.
For ME4(5\AA{}), a similar trend emerges.
Up to a 0.2~kJ/mol energy threshold, the MAE and MAX remain unchanged, while the ME decreases from 0.2 to -0.1~kJ/mol.
Here, between none and all tetramers are excluded for the different X23 systems, the average being $\sim$66~\%.
For both methods, at higher cutoffs, then up to all of the respective multimers are excluded, leading to the results of the respective one-order-lower method, ME2(4\AA{}) and ME3(5\AA{}).
Overall, at the trimer level, it seems that it is not possible to save a significant number of trimer calculations without sacrificing accuracy using energy-based cutoffs.
On the tetramer level, on the other hand, an energy interaction threshold of 0.1 or 0.2~kJ/mol can significantly reduce the number of high-level tetramer calculations to be carried out, without experiencing significant accuracy losses.

\subsection{Trimer Interactions for Forces and Stress}

Having discussed lattice energies in great depth, we now shift our focus to forces, stress tensors, and consequently, lattice relaxations.
Previously\cite{Hoja2023}, we discussed the application of our embedding scheme to forces and stress for up to dimers.
We have shown that, even though monomer and dimer embedding already significantly reduce the deviation of the force and stress tensor components to those of PBE0+MBD compared to PBE+MBD, due to the often flat nature of the potential energy surfaces of the X23 systems, the obtained cell volumes were still lacking.
To establish an initial assessment of the impact of trimer interactions, we had performed single-point energy calculations at various cell volumes of the ammonia crystal, showing that the utilization of trimer embedding shifts the minimum-energy volume closer to that of PBE0+MBD by a great extent, compared to dimer embedding (Fig. 2 of Ref. \citenum{Hoja2023}).
We are now in a position to explicitly look at the forces, stress tensor, and cell volumes obtained through the utilization of trimer embedding.

\begin{table}
\setlength{\tabcolsep}{2pt}
\caption{\label{tbl:forces} Mean absolute errors (MAE) and maximum errors (MAX) for the computed atomic force components and non-zero stress tensor components of the X23 set in comparison to PBE0+MBD results, calculated at PBE+MBD-optimized structures.}
\begin{tabular}{@{\extracolsep{4pt}}lrrrr@{}}
\hline\hline
 &\multicolumn{2}{c}{Forces (eV/\AA{})} & \multicolumn{2}{c}{Stress (eV/\AA{}$^3)$} \\  \cline{2-3}\cline{4-5}
Method &       MAE & MAX  &  MAE & MAX\\
\hline
PBE+MBD              &  0.226 &  1.422 & 0.01887 & 0.04963 \\
ME1                  &  0.023 &  0.216 & 0.00073 & 0.00299 \\
ME2(4\AA{})          &  0.006 &  0.055 & 0.00057 & 0.00228 \\
ME3(3\AA{})          &  0.007 &  0.107 & 0.00031 & 0.00150 \\
ME3(4\AA{})          &  0.004 &  0.053 & 0.00014 & 0.00071 \\
ME3(5\AA{})          &  0.004 &  0.045 & 0.00022 & 0.00136 \\
ME3(6\AA{})          &  0.007 &  0.078 & 0.00044 & 0.00228 \\
\hline\hline
\end{tabular}\\
\end{table}

First, we have calculated the trimer embedding forces and stress tensors on top of PBE+MBD-optimized structures of the X23 set with light settings, utilizing various cutoff distances.
The resulting errors, together with a selection of earlier investigated methods\cite{Hoja2023} to offer perspective, are shown in Table \ref{tbl:forces}.
While PBE+MBD had a mean average error (MAE) of force components of 0.226~eV/\AA{} compared to PBE0, ME1 could reduce this error by a factor of ten and ME2(4\AA{}) even further reduced the MAE to 0.006~eV/\AA{} while having a maximum error (MAX) of 0.055~eV/\AA{}.
Going now to trimers, there is little improvement to be seen.
ME3(3\AA{}) performs slightly worse than the best dimer embedding method, while ME3(4\AA{}) and ME3(5\AA{}) show marginal improvement, both having a MAE of 0.004~eV/\AA{} and a MAX of 0.053 and 0.045~eV/\AA{}, respectively.
Going to a 6~\AA{} cutoff increases the error again, showing similar behavior as the trimer embedding for energies, where a certain degree of error cancellation yielded the best result for smaller cutoff distances.
However, as even with dimer embedding the errors were already almost within our optimization convergence criterion, the small improvement achievable through the inclusion of trimer interactions is more than sufficient.

In comparison, the effect of trimer interactions on the stress tensor is much more pronounced.
For all investigated cutoff distances, trimer embedding shows an improvement compared to dimer embedding, the best method being ME3(4\AA{}), which reduces the MAE of the non-zero stress components by a factor of about four to 0.00014~eV/\AA{}$^3$, compared to ME2(4\AA{}).

\begin{figure*}[!]
\includegraphics[width=\textwidth]{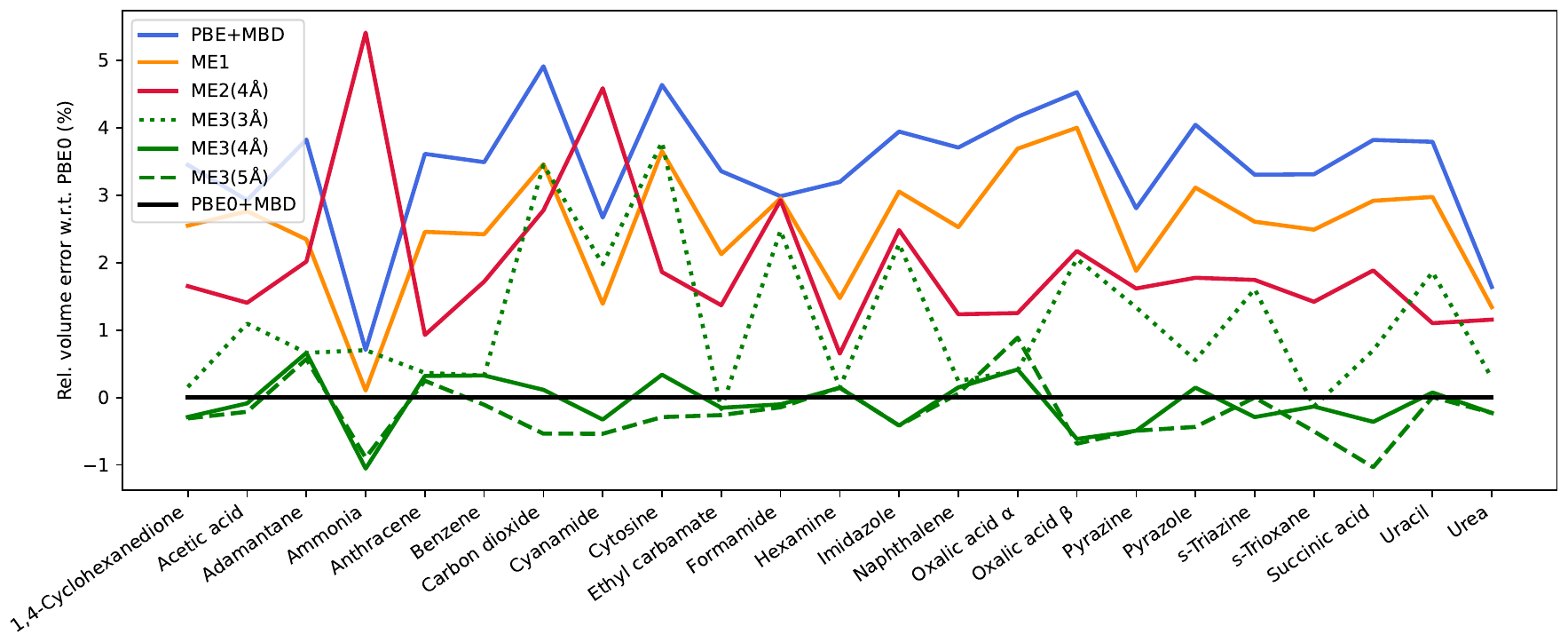}
\caption{\label{fig:volumes} Relative volume errors for the optimized X23 cells with different levels of multimer embedding compared to the PBE0+MBD optimized cell volumes in \%.}
\end{figure*}

\begin{table}
\setlength{\tabcolsep}{2pt}
\caption{\label{tbl:opt} Errors of the cell volumes (in \%) and corresponding lattice energies (in kJ/mol) of the X23 set obtained from lattice relaxations of different embedding methods using light settings, compared to PBE0+MBD cell volumes and lattice energies.}
\begin{tabular}{@{\extracolsep{4pt}}lrrrrrr@{}}
\hline\hline
 &\multicolumn{3}{c}{$V$ (\%)} & \multicolumn{3}{c}{$E_{\rm latt}$ (kJ/mol)} \\  \cline{2-4}\cline{5-7}
Method &    MRE & MARE & RMAX &   ME &  MAE & MAX\\
\hline
PBE+MBD           &  3.4 &  3.4 &  4.9 &  1.6 &  2.3 &  9.4 \\
ME1               &  2.5 &  2.5 &  4.0 &  3.2 &  3.7 &  9.9 \\
ME2(4\AA{})       &  2.0 &  2.0 &  5.4 &  3.4 &  3.4 &  5.9 \\
ME3(3\AA{})       &  1.1 &  1.2 &  3.8 &  1.1 &  1.2 &  4.4 \\
ME3(4\AA{})       & -0.1 &  0.3 &  1.1 & -0.2 &  0.4 &  1.1 \\
ME3(5\AA{})       & -0.2 &  0.4 &  1.0 & -0.5 &  0.6 &  1.4 \\
\hline\hline
\end{tabular}\\
\end{table}

Next, we have performed lattice relaxations utilizing trimer embedding at 3, 4, and 5~\AA{} with light settings and compared the resulting cell volumes to those of PBE0+MBD.
Fig. \ref{fig:volumes} shows the resulting volume differences, which are then summarized in Table \ref{tbl:opt}.
It can be seen that ME3(4\AA{}) and ME3(5\AA{}) are able to reproduce the PBE+MBD cell volumes significantly better than the previously studied lower-order multimer embedding methods.
ME3(4\AA{}) shows a mean relative error (MRE) of -0.1~\%, a mean absolute relative error (MARE) of 0.3~\%, and a maximum relative error (RMAX) of 1.1~\%.
For ME3(5\AA{}), the MRE, MARE, and RMAX are very similar, being -0.2~\%, 0.4~\%, and 1.0~\%, respectively.
Overall, both methods provide an accurate approximation of the stress tensor, leading to excellent cell volumes.
ME3(3\AA{}), on the other hand, even though it still shows an improvement on average, lacks crucial trimer interactions for roughly a third of the X23 systems, as can be seen in Fig \ref{fig:volumes}, resulting in similar or worse cell volumes than ME1 or ME2 in those cases.
Comparing the ME3 lattice energies at their optimized geometries (also Table \ref{tbl:opt}) with those at the PBE0+MBD-optimized geometries from our previous publication, there is now essentially no difference, due to the cell volumes being so similar and the flat nature of the potential energy surfaces.

\subsection{Trimer Interactions for Harmonic Vibrational Properties}

\begin{table}
\setlength{\tabcolsep}{5pt}
\caption{\label{tbl:vib} Mean errors (ME) and mean absolute errors (MAE) of the gamma-point vibrational/phonon frequencies of the X23 set in cm$^{-1}$ and vibrational free energies at 300~K normalized per molecule in kJ/mol (converged q-grid) for different embedding methods, calculated at the respective optimized geometries and compared to PBE0+MBD results (light settings).}
\begin{tabular}{@{\extracolsep{2pt}}lrrr@{}}
\hline\hline
 Quantity       & Method      &   ME &  MAE \\
\hline
& PBE+MBD              &  -47.8 &   48.7 \\
& ME1                  &   -0.8 &    3.9 \\
$\nu(\Gamma)$ & ME2(4\AA{})          &   -0.6 &    2.4 \\
& ME3(3\AA{})          &    0.3 &    2.5 \\
& ME3(4\AA{})          &    0.0 &    1.3 \\
& ME3(5\AA{})          &    0.1 &    1.8 \\
\hline
& PBE+MBD              &  -10.7 &   10.7 \\
& ME1                  &   -1.0 &    1.0 \\
$F_{\rm vib}^a$ & ME2(4\AA{})          &   -0.6 &    0.7 \\
& ME3(3\AA{})          &   -0.3 &    0.6 \\
& ME3(4\AA{})          &   -0.2 &    0.3 \\
& ME3(5\AA{})          &   -0.1 &    0.4 \\
\hline\hline
\end{tabular}\\
$^a$ evaluated at 300~K and normalized per molecule
\end{table}

Finally, we discuss the impact of the inclusion of trimer interactions on phonon calculations.
We therefore have calculated the gamma point frequencies as well as vibrational free energies for the X23 structures using light species default settings for ME3(3\AA{}), ME3(4\AA{}), and ME3(5\AA{}) at their respective optimized geometries.
The results can be seen in Table \ref{tbl:vib}.
Previously\cite{Hoja2023}, we have shown that ME1 already makes up $\sim$90~\% of the difference between PBE+MBD and PBE0+MBD, owing to the fact that a large proportion of molecular crystal vibrations are intramolecular.
Going to dimer embedding leads to further improvements, even though they are much less pronounced.
Considering trimers now, we observe yet again small improvements.
In the case of ME3(3\AA{}), the ME of the gamma point frequencies is 0.3~cm$^{-1}$, while the MAE with 2.5~cm$^{-1}$ remains largely unchanged compared to ME2(4\AA{}).
The ME of ME3(4\AA{}) is virtually zero, and its MAE is 1.3~cm$^{-1}$,
the lowest of all methods investigated.
For ME3(5\AA{}), the MAE then increases back to 1.8~cm$^{-1}$.

The vibrational free energies have been calculated at converged q-grids, a temperature of 300~K, and have been normalized per molecule.
Here, all trimer embedding methods have lower errors than ME2(4\AA{}).
ME3(3\AA{}) has a ME of -0.3~kJ/mol and a MAE of 0.6~kJ/mol, which are further decreased to -0.2~kJ/mol and 0.3~kJ/mol for ME3(4\AA{}), respectively. 
For ME3(5\AA{}) the ME and MAE amount to -0.1~kJ/mol and 0.4~kJ/mol, respectively.

In the supplementary material, we also provide an excerpt of the resulting phonon density of states of uracil using different embedding orders.
Fig. S2 shows the phonon density of states in the frequency range of 400-610~cm$^{-1}$, and Fig. S3 shows the low-frequency range of up to 160~cm$^{-1}$.
In the first case, we can see that all embedding methods show good agreement with PBE0+MBD, while for PBE+MBD all frequencies are shifted by approx. -20~cm$^{-1}$.
This is due to the fact that in this frequency range we already only observe internal vibrational modes.
In the second case, as for the low-frequency modes the cell volume plays a much larger role, only ME3(4\AA{}) is able to reproduce the PBE+MBD's phonon density of states' peak location and shape.

Note that previously we have also presented the results of phonon calculations at internally relaxed geometries with PBE0+MBD optimized cells; however, as ME3(4\AA{}) and ME3(5\AA{}) are able to reproduce the PBE0+MBD cell volumes very well, the differences are much less pronounced.
To give an example, while ME3(3\AA{}) has a gamma point frequency MAE of 2.7~cm$^{-1}$ at the optimized cell and 1.6~cm$^{-1}$ at the PBE0 cell, for ME3(4\AA{}) the respective values are 1.5~cm$^{-1}$ and 1.2~cm$^{-1}$, only a 0.3~cm$^{-1}$ difference.

\section{Conclusion}
We have extended our previously introduced\cite{Hoja2023} subtractive embedding scheme for molecular crystals, which can now include up to tetramer interactions for energies and up to trimer interactions for forces, stress, and harmonic vibrational properties.
This embedding scheme serves to reduce the computational cost of molecular crystal calculations, as only multimers up to the highest included multimer order must be calculated with the high-level method (hybrid density functional), rather than the fully periodic system.
Herein, we have presented the effects of the additionally included multimer orders by approximating periodic PBE0+MBD results through the embedding of PBE0+MBD multimers into periodic PBE+MBD calculations for the X23 benchmark set of molecular crystals.

For the approximation of lattice energies, we have shown that tetramer interactions do not play as a significant role as trimer interactions.
However, we have found that at the tetramer level there is much less reliance on error cancellation, leading to smaller errors even at larger multimer cutoff distances.
This follows the expected behavior of the multimer embedding scheme eventually converging to the periodic high-level result when utilizing increasingly high multimer orders and cutoff distances.
Nevertheless, it appears that corrections up to the trimer level are sufficient for the accurate representation of lattice energies in this case. 

Furthermore, we have investigated the possibility of including tri- and tetramer types, where not all monomers are within cutoff distance of each other, as well as applying different cutoff distances for each multimer order.
In both cases, we concluded that there was no clear benefit from employing such strategies.
Moreover, in an attempt to increase the efficiency of our embedding scheme, we have introduced energy-based cutoffs for multimer selection, where high-level multimer calculations are only carried out if the low-level interaction energy exceeds a certain threshold.
Here, we could show that energy thresholds of up to 0.2~kJ/mol do not significantly impact the quality of the embedding result, while saving an average of 10~\% of high-level trimer calculations for ME3(4\AA{}) and 66~\% of high-level tetramer calculations for ME4(5\AA{}).

While atomic forces were already well described at the dimer level, the inclusion of trimer interactions significantly improved the description of the stress tensor, resulting in much better cell volumes for lattice relaxations.
It is now possible to reproduce the PBE0+MBD cell volumes with a mean absolute relative error (MARE) of only 0.4~\%, with no individual error being larger than 1.1~\%.
For comparison, dimer embedding is only able to reach a MARE of cell volumes of 2~\%.

Finally, regarding harmonic vibrational properties, trimer embedding also provides an additional small improvement, although gamma-point frequencies and vibrational free energies were already well described at the monomer and dimer level.
However, as only at the trimer level satisfactory cell volumes can be obtained, we also see a corresponding improvement in the resulting phonon density of states for low frequencies, which can be crucial when studying for instance THz spectra of molecular crystals.

In summary, we have now thoroughly investigated our multimer embedding scheme for the PBE0+MBD:PBE+MBD case. 
By extending the lattice energy approximation to tetramer interactions, we were able to study in detail the convergence behavior of our method w.r.t. the included multimers and cutoffs, leading to a much more complete understanding of our method.
For embedding PBE0+MBD into PBE+MBD the ME3(4\AA{}) level provides the best compromise between accuracy and computational cost, yielding excellent embedding results for lattice energies, cell volumes, as well as harmonic vibrational properties.

\section*{Supplementary Material}
The supplementary material contains additional tables and figures with multimer numbers for different cutoff distances, lattice energies and respective statistics, individual unit cell volumes, and phonon densities of states.

\begin{acknowledgments}
The computational results have been achieved in part using the Austrian Scientific Computing (ASC) infrastructure.
\end{acknowledgments}

\section*{Data Availability Statement}
The data that supports the findings of this study are openly available in a Zenodo repository\cite{SD} at \href{https://doi.org/10.5281/zenodo.17975628}{https://doi.org/10.5281/zenodo.17975628}, within the article, and its supplementary material.

\end{document}